\documentclass[a4paper,conference]{IEEEtran}

\usepackage[utf8]{inputenc}
\usepackage[T1]{fontenc}
\usepackage{url}
\usepackage{ifthen}
\usepackage{cite}
\usepackage[cmex10]{amsmath} 
\usepackage[export]{adjustbox}
\usepackage{multirow}
\usepackage{rotating}
\usepackage{booktabs}
\usepackage{caption}
\begin{document}
\title{Near Maximum Likelihood Decoding\\with Deep Learning}

\author{%
  \IEEEauthorblockN{Eliya Nachmani, Yaron Bachar, Elad Marciano, David Burshtein and Yair Be'ery}
  \IEEEauthorblockA{School of Electrical Engineering, Tel-Aviv University, Israel\\
Emails: enk100@gmail.com, yaronbac@gmail.com, eladmarc@gmail.com, burstyn@eng.tau.ac.il, ybeery@eng.tau.ac.il}
}



\maketitle

\begin{abstract}
A novel and efficient neural decoder algorithm is proposed. The proposed decoder is based on the neural Belief Propagation algorithm and the Automorphism Group. By combining neural belief propagation with permutations from the Automorphism Group we achieve near maximum likelihood performance for High Density Parity Check codes. Moreover, the proposed decoder significantly improves the decoding complexity, compared to our earlier work on the topic. We also investigate the training process and show how it can be accelerated. Simulations of the hessian and the condition number show why the learning process is accelerated. We demonstrate the decoding algorithm for various linear block codes of length up to 63 bits.
\end{abstract}

\section{INTRODUCTION}
In the last few years Deep Learning methods were applied to communication systems, for example in \cite{farsad2017detection,AutoencoderComm,samuel2017deep,liang2017iterative,dorner2017deep,Shengliang, power_ofdm,Haowen_Lin}. Furthermore,
Deep Neural decoders is a new approach for decoding linear block codes. In \cite{nachmani,nachmani2017rnn,lugosch,nachmani2017deep} it has been shown that deep neural decoders can improve the existing belief propagation methods for decoding high density parity check codes (HDPCs). Other methods for using deep learning to decode error correcting codes were proposed in \cite{tenbrink, cammerer2017scaling, quantumCodes}. In this work we combine the deep recurrent neural decoder of \cite{nachmani2017rnn} with permutations from the Automorphism Group as defined in \cite{pless1978fj}. The combined architecture is defined by $I_{permutations}$ blocks, each of which contains $I_{BP}$ iterations of neural belief propagation followed by permutation. We show that this architecture achieves near maximum-likelihood performance for various BCH codes of up to $63$ bits long with significantly lower complexity then the corresponding mRRD decoder \cite{dimnik2009improved}. We also investigate the training process of the deep neural decoder and show how the learning can be accelerated by adding penalties to the loss function. We argue that this penalties transform the manifold of the loss function into an isotropic manifold which is easy to optimize. Simulations of the Hessian matrix of the loss function support this claim.

\section{THE NEURAL BELIEF PROPAGATION ALGORITHM} \label{basicrnn}
We start with a brief description of the deep neural network proposed in \cite{nachmani,nachmani2017rnn}. The deep neural decoder is a message passing algorithm  parameterized as a deep neural network. 
The input to the neural network is the set of LLR values, $v=1,2,\ldots,N$,
$$
l_v = \log\frac{\Pr\left(C_v=1 | y_v\right)}{\Pr\left(C_v=0 | y_v\right)}
$$
where $N$ is the block length of the code, $y_v$ is the channel output corresponding to the $v$th codebit. The neural decoder consists of pairs of odd and even layers. For odd $i$ layer, 
\begin{align}
&x_{i,e=(v,c)} =\nonumber \\ 
&=\tanh \Biggl( \frac{1}{2} \Biggr.\Biggl(l_v + \sum_{e'=(c',v),\: c'\ne c} w_{e,e'} x_{i-1,e'}\Biggr) \Biggl. \Biggr ) 
\label{eq:x_ie_RB_NN}
\end{align}
for even $i$ layer,
\begin{equation}
x_{i,e=(c,v)} = 2\tanh^{-1} \left( \prod_{e'=(v',c),\: v'\ne v}{x_{i-1,e'}}\right)
\label{eq:x_ie_LB_NN}
\end{equation}
and for output layer,
\begin{equation}
o_v = \sigma \left( l_v + \sum_{e'=(c',v)} \tilde{w}_{v,e’} x_{2L,e'} \right)
\label{eq:ov_NN}
\end{equation}
where $\sigma(x) \equiv \left( 1+e^{-x} \right)^{-1}$ is a sigmoid function. Please note that equations \eqref{eq:x_ie_RB_NN},\eqref{eq:x_ie_LB_NN} define recurrent neural network, as the learnable weights $w_{e,e'}, \tilde{w}_{v,e’}$ are tied.

\section{THE PROPOSED DEEP NEURAL NETWORK DECODER}

\subsection{Architecture}
\begin{figure*}
\centering
\includegraphics[width=\textwidth]{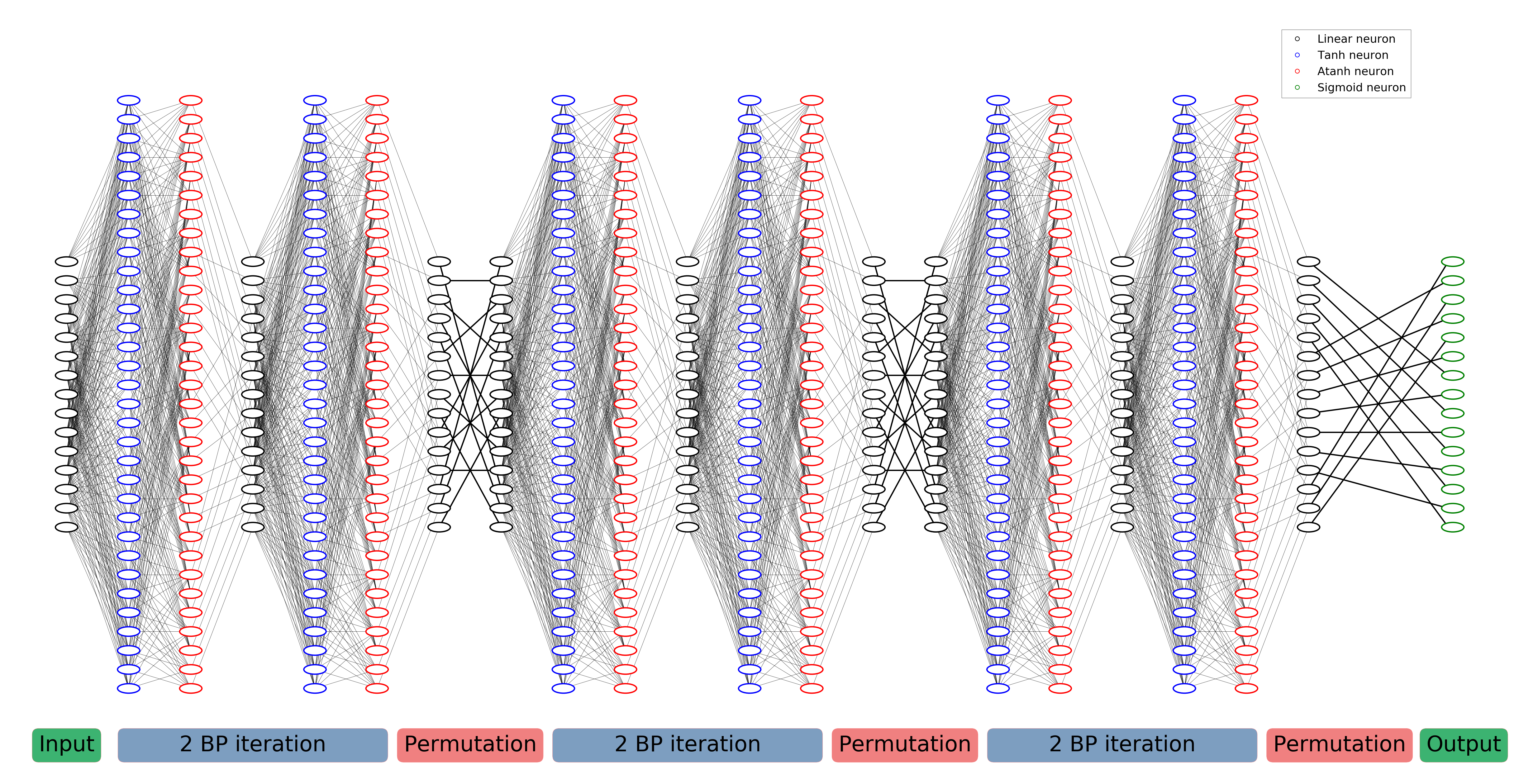}
\caption{Deep Neural Network Architecture For BCH(15,11) with 3 permutations and 2 belief propagation iterations for each permutation. The permutations have bold lines. The self message $o^j_{i,v}$ was removed from the diagram for a cleaner view.} 
	\label{fig:DNN_fig}
\end{figure*}

The proposed neural network is composed of $ I_{permutations} $ blocks. Each block contains $I_{BP}$ layers of neural belief propagation, which are described below. Between each two successive blocks, we apply a permutation from the automorphism group. Lastly, we apply the corresponding inverse permutation, to obtain the decoded codeword. The proposed architecture is illustrated in Figure \ref{fig:DNN_fig}.

We re-parameterized the deep neural network decoder from section \ref{basicrnn}. In the $j$-th block, $I_{BP}$ iterations of neural belief-propagation are performed as follows:

For each variable node in the $i$-th layer,
\begin{align}
x^j_{i,e=(v,c)} = \tanh  \left( \frac{1}{2} ( o^j_{i-1,v} - x^j_{i-1,e=(c,v)} )  \right) 
\label{eq:x_ie_RB_NN_new}
\end{align}

For each check node in the $i$-th layer,
\begin{equation}
x^j_{i,e=(c,v)} = 2\tanh^{-1} \left( \prod_{e'=(v',c),\: v'\ne v}{x^j_{i-1,e'}}\right)
\label{eq:x_ie_LB_NN_new}
\end{equation}

For mid-output node in the $i$-th layer, 
\begin{equation}
o^j_{i,v} =  o^{j}_{0,v} + \sum_{e'=(c',v)} w_{e'} x^j_{i,e'} 
\label{eq:ov_NN_new}
\end{equation}

The output of the j-th block:
\begin{equation}
c^j_v = \pi_j(o^j_{i=I_{BP},v})
\end{equation}

The initialization:
\begin{align}
o^j_{0,v} = 
\left\{\begin{matrix}
 l_v &,  j=0 \\ 
 c^{j-1}_v &,  j>0
\end{matrix}\right.
\label{eq:o_init_new}
\end{align}
\begin{align}
x^j_{0,e} = 0
\label{eq:x_init_new}
\end{align}

After each BP block, an appropriate inverse permutation is applied:
\begin{equation}
d_v^j = (\pi_1^{-1} \cdot \pi_2^{-1} \cdot ... \cdot \pi_j^{-1})c_v^j
\end{equation}

Note that the neural weights $w_{e'}$ are tied along the neural network graph. Also note, that the new parametrization of the neural belief propagation decoder was easier to optimize, and converged faster to a better performance. 

As in \cite{nachmani,nachmani2017rnn,lugosch,nachmani2017deep} 
the proposed architecture preserves the symmetry conditions, therefore we can train the neural network with noisy versions of a single codeword.

Also note, that in order to be consistent with the BP algorithm, one needs to multiply $x_{i-1,e=(c,v)}^j$ in \eqref{eq:x_ie_RB_NN_new} by $w_e$. However, this multiplication did not have any significant influence on the results obtained.

\subsection{Loss function}
The loss of the neural network is composed of three constituents:
\begin{itemize}


\item A multi-loss cross-entropy between $\tilde{d}^j_v \equiv \sigma(d^j_v)$ and the correct codeword $y_v$. This is a loss term concerned with the output of the BP blocks:

\begin{equation}
L_1^j=-\frac{1}{N}\sum_{v=1}^{N}y_{v}\log(\tilde{d}^j_v)+(1-y_{v})\log(1-\tilde{d}^j_v)
\label{eq:cross_entropy2}
\end{equation}

\item A sub multi-loss cross-entropy between $\tilde o^j_{i,v} \equiv \sigma(o^j_{i,v})$ and the correct codeword $y_v$. This term involves inner-BP marginalizations:

\begin{equation}
L_2^{j,i}=-\frac{1}{N}\sum_{v=1}^{N}y_{v}\log( o^j_{i,v})+(1-y_{v})\log(1- o^j_{i,v})
\label{eq:cross_entropy2}
\end{equation}

\item $l_2$-norm of the weights $w_{v}$, $w_{v,e'}$:
\begin{equation}
L_3=\sum_{v} \left \| w_{v} \right \|^2 + \sum_{v,e}\left \| w_{v,e} \right \|^2 
\label{eq:cross_entropy3}
\end{equation}

\end{itemize}

The total loss is:
\begin{equation}
L = \sum_{j} (L_1^j + \lambda \cdot L_3) + \sum_{j,i}L_2^{j,i} 
\label{eq:tot_loss}
\end{equation}

\section{EXPERIMENTS}
\subsection{Neural Network Training and Dataset} 
We implemented the proposed neural network in TensorFlow framework. The neural network was optimized with RMSPROP~\cite{rmsprop}. As in \cite{nachmani,nachmani2017rnn,lugosch,nachmani2017deep}, the dataset consisted of the zero codeword and an AWGN channel. We used the cycle reduced parity check matrix from \cite{cycle_reduce}. Due to large number of layers in our network, and the fact that the network is a recurrent neural network, gradient clipping was applied to avoid gradient exploding throughout the learning process. Clipping threshold of $c_{grad}=0.1$ was used. The $l_2$-Loss term was added with a factor of $\lambda$.  Note that we use the three terms $L_1, L_2, L_3$ of the loss for training. The weights were constrained to have non-negative values. In all of our experiments no overfitting was observed.

The architecture was tested on BCH codes. Their automorphism group is described in detail in \cite{automorphism}. The permutations were chosen randomly using the product-replacement algorithm \cite{cycle_reduce}, which has the $N_{pr}$ and $K_{pr}$ parameters. $N_{pr}$ is the size of the group of permutations the algorithm builds, and $K_{pr}$ is the initial number of iterations, used to build this permutations-reservoir. In Table \ref{table:params} we provide details about the parameters configurations of the network.

\subsection{BCH(63,45)}
Batch size was set to $160$, with $20$ examples per SNR. The SNR varied from $1dB$ to $8dB$ in the training process, and from $1dB$ to $5dB$ in the validation process. 
The neural network comprises $ I_{permutations} = 50$ permutations, and each block contains $ I_{BP} =2 $ BP iterations. A total of $100$ BP iterations correspond to a deep neural network with $200$ layers.
\subsection{BCH(63,36)}
Batch size was set to $120$, with $30$ examples per SNR. The SNR varied from $1dB$ to $6dB$ in the training process, and from $3dB$ to $4.5dB$ in the validation process. 
The neural network comprised $ I_{permutations} = 300$ permutation, and each block contains $ I_{BP} =2 $ belief propagation iteration. This configuration represents $600$ Belief Propagation iterations which correspond to deep neural network with $1200$ layers.

\begin{center}
\begin{tabular}{ |c|c|c|c| } 
\hline
 & Parameter & BCH(63,45) & BCH(63,36)\\
\hline
\multirow{2}{1em}{\begin{turn}{90}BP\end{turn}
} & $I_{BP}$ & 2 & 2 \\ 
& llr clip & 15 & 15 \\ 
\hline
\hline
\multirow{3}{1em}{\begin{turn}{90}RRD\end{turn}
} & $I_{Permutatoins}$ & 50 & 300 \\ 
& $N_{pr}$ & 20 & 1000 \\ 
& $K_{pr}$  & 60 & 4000 \\
\hline
\hline
\multirow{8}{1em}{\begin{turn}{90}Neural Network\end{turn}
} & learning rate & 1e-3 & 1e-3 \\ 
& batch size & 160 & 120 \\ 
& batch size / snr & 20 & 30 \\ 
& SNR range & 1-8dB & 1-6dB \\
& $\lambda$ & 100 & $10^{12}$ \\
& gradient clipping & 0.1 & 0.1 \\
& network depth & 200  & 1200 \\
\hline
\end{tabular}
\captionof{table}{Parameter Configuration of the Model}
\label{table:params}
\end{center}

\subsection{Results}
In the following figures, "Perm-RNN-i-j-k" denotes our proposed decoder,  with i parallel branches, j permutations and k BP iterations between two consecutive permutations; "mRRD-i" denotes the classical mRRD decoder with i branches; and "mRRD-RNN-i-j-k" denotes the mRRD-RNN decoder with i branches, j blocks of BP, each with k iterations.

\begin{figure}[thpb]
	\centering	\includegraphics[width=1.1\linewidth]{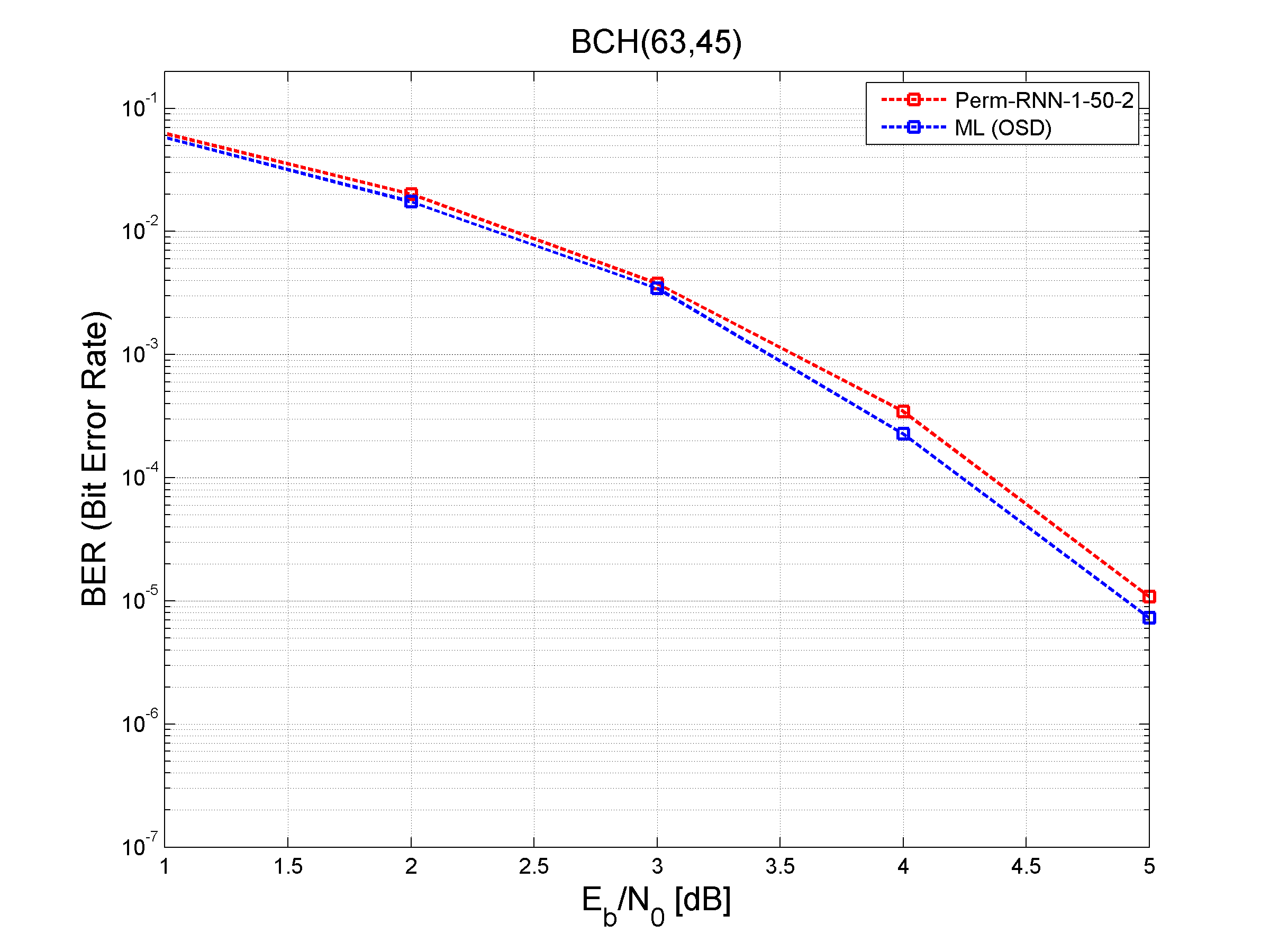}
	\caption{BER results for BCH(63,45) code}
	\label{fig:bch_63_45_ber_permRNN}
\end{figure}

\begin{figure}[thpb]
	\centering	\includegraphics[width=1.1\linewidth]{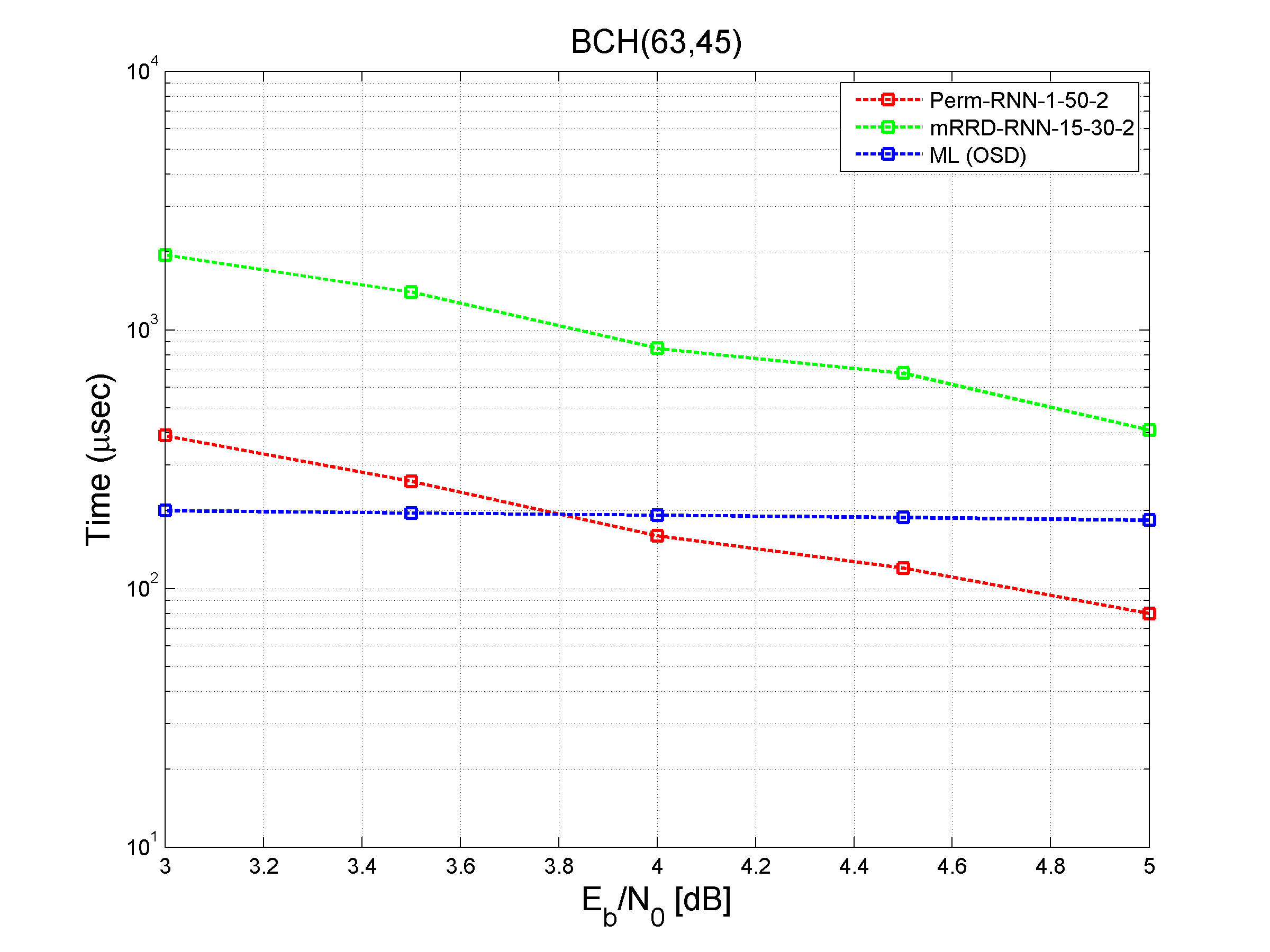}
	\caption{Running time comparison for BCH(63,45) code}
	\label{fig:bch_63_45_time_permRNN}
\end{figure}

\begin{figure}[thpb]
	\centering	\includegraphics[width=1.1\linewidth]{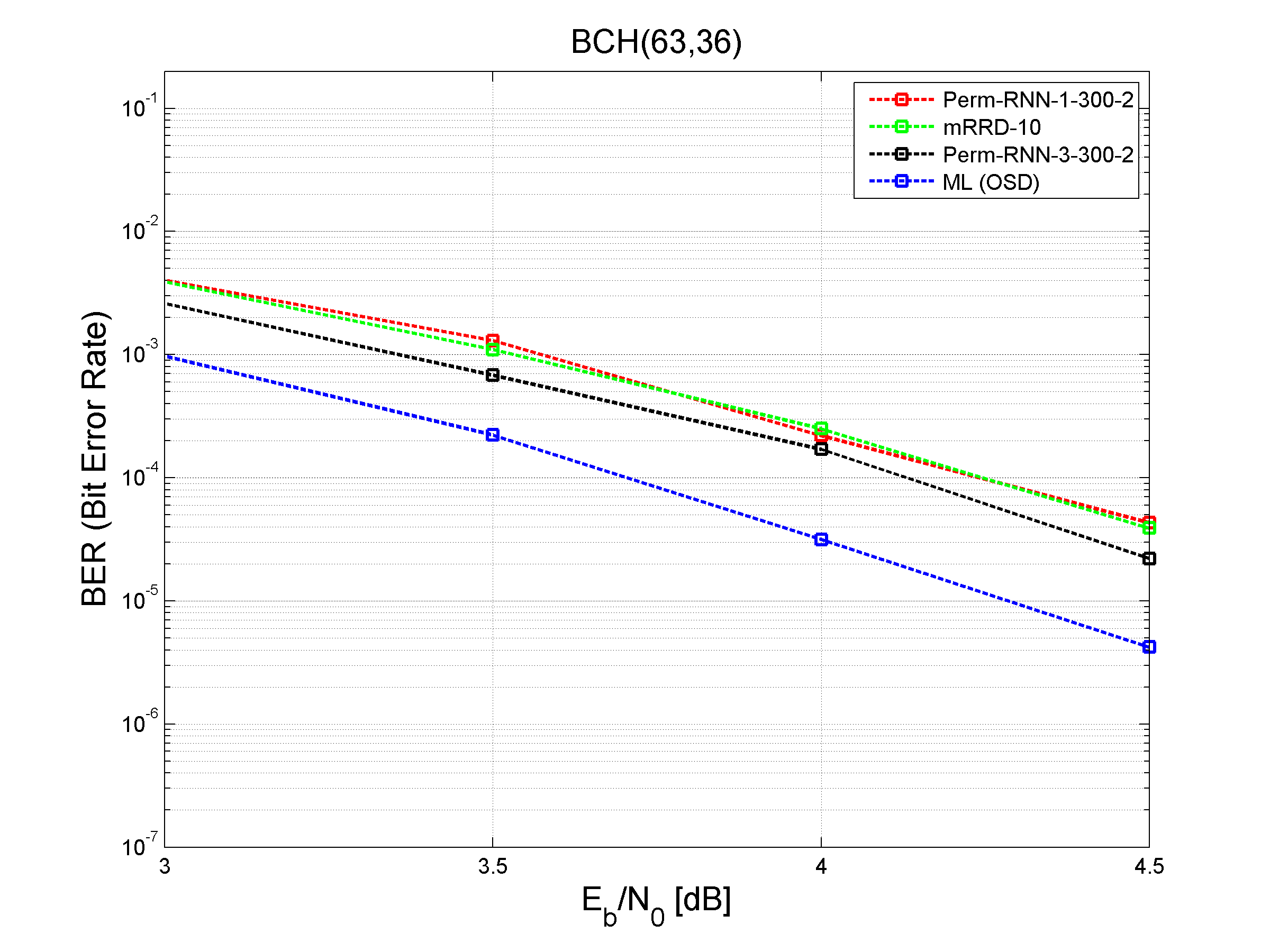}
	\caption{BER results for BCH(63,36) code}
	\label{fig:bch_63_36_ber_permRNN}
\end{figure}

\begin{figure}[thpb]
	\centering	\includegraphics[width=1.1\linewidth]{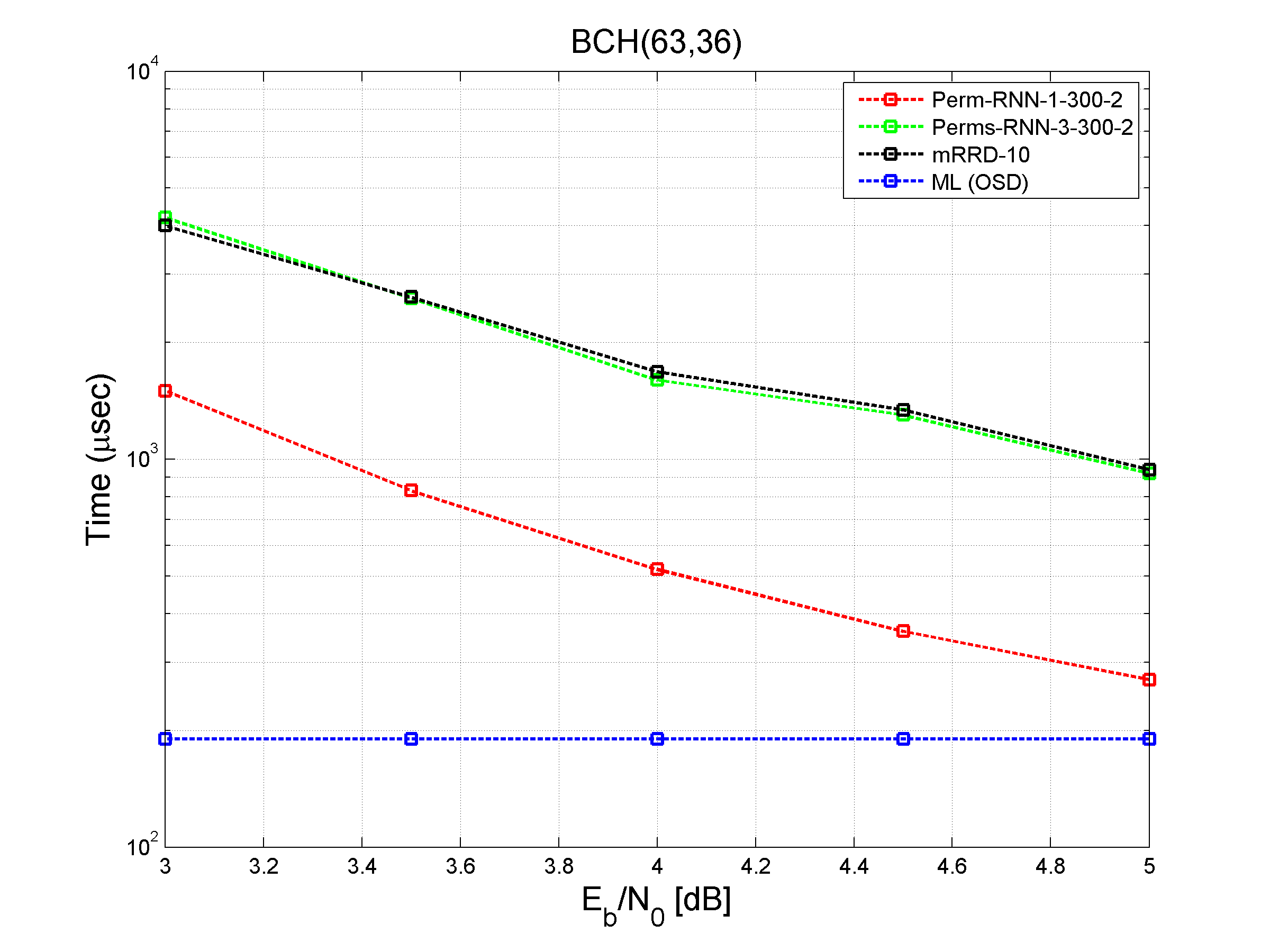}
	\caption{Running time comparison for BCH(63,36) code}
	\label{fig:bch_63_36_time_permRNN}
\end{figure}

In Figure \ref{fig:bch_63_45_ber_permRNN} we provide the bit-error-rate for $BCH(63,45)$ code for our proposed decoder. The maximum-likelihood estimate was obtained by the OSD algorithm \cite{fossOSD}. We observe near maximum likelihood performance with our proposed decoder, with a gap of up to 0.2dB to ML. The runtime of the proposed neural decoder is lower than OSD's when SNR is bigger than 3.8dB, as shown in figure \ref{fig:bch_63_45_time_permRNN}. In Figures \ref{fig:bch_63_36_ber_permRNN} and \ref{fig:bch_63_36_time_permRNN} we provide the bit-error-rate and the running time for $BCH(63,36)$ code for our proposed decoder. The maximum-likelihood estimate was obtained by the 2nd order OSD algorithm \cite{fossOSD}, and the mRRD performance was obtained using 10-parallel mRRD decoder \cite{dimnik2009improved}. We have a gap of 0.25-0.5dB to achieve maximum likelihood performance with our proposed decoder.

Note, that the overall decoding time of our decoder is substantially smaller than the mRRD's decoding time for the (63,36) code, with a factor of up to 3.5. In addition, only one neural decoder was needed to match the performance 10-parallel mRRD decoder. Also note, that OSD's main disadvantage of parallel implementation is not encountered in the neural decoder.

\begin{figure}[thpb]
	\centering	\includegraphics[width=1.1\linewidth]{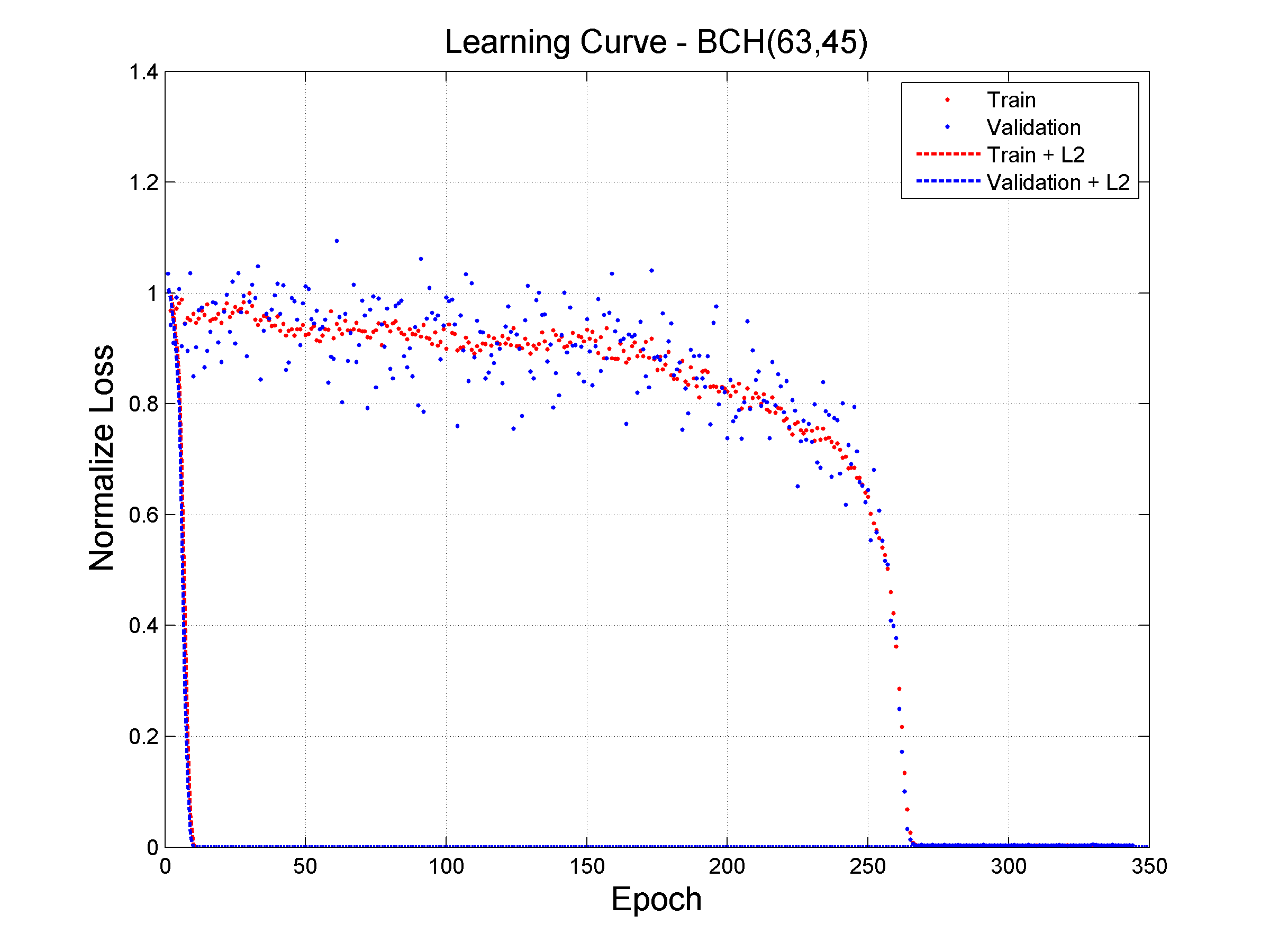}
	\caption{Learning Curve for BCH(63,45) code}
	\label{fig:bch_63_45_Learning_curve}
\end{figure}

In Figure \ref{fig:bch_63_45_Learning_curve} we provide the learning curve for $BCH(63,45)$ code. The learning rate was constant during the training process, yet the loss significantly drops at some stage of the training. For training without $l_2$-norm, the drop occurs in epoch $265$, and most of the improvement occurs at the same time. Training with $l_2$-norm accelerated the learning process: the loss dropped at epoch $8$, as if the training process was accelerated by factor $33$. We will investigate and discuss the dropping phenomenon and the $l_2$-acceleration at section \ref{sssec:hess}.

\section{TRAINING ACCELERATION}\label{sssec:hess}
As introduced in the previous section, during the training the loss drops significantly. This phenomenon usually occurs while training very deep neural networks. Note that our proposed network for $BCH(63,45)$ contained $200$ layers. As shown in \cite{dauphin2014identifying}, this phenomenon can be explained by the existence of saddle points in the loss-surface of the network.  
\subsection{Hessian simulation}
\begin{figure}[thpb]
	\centering	\includegraphics[width=1.1\linewidth]{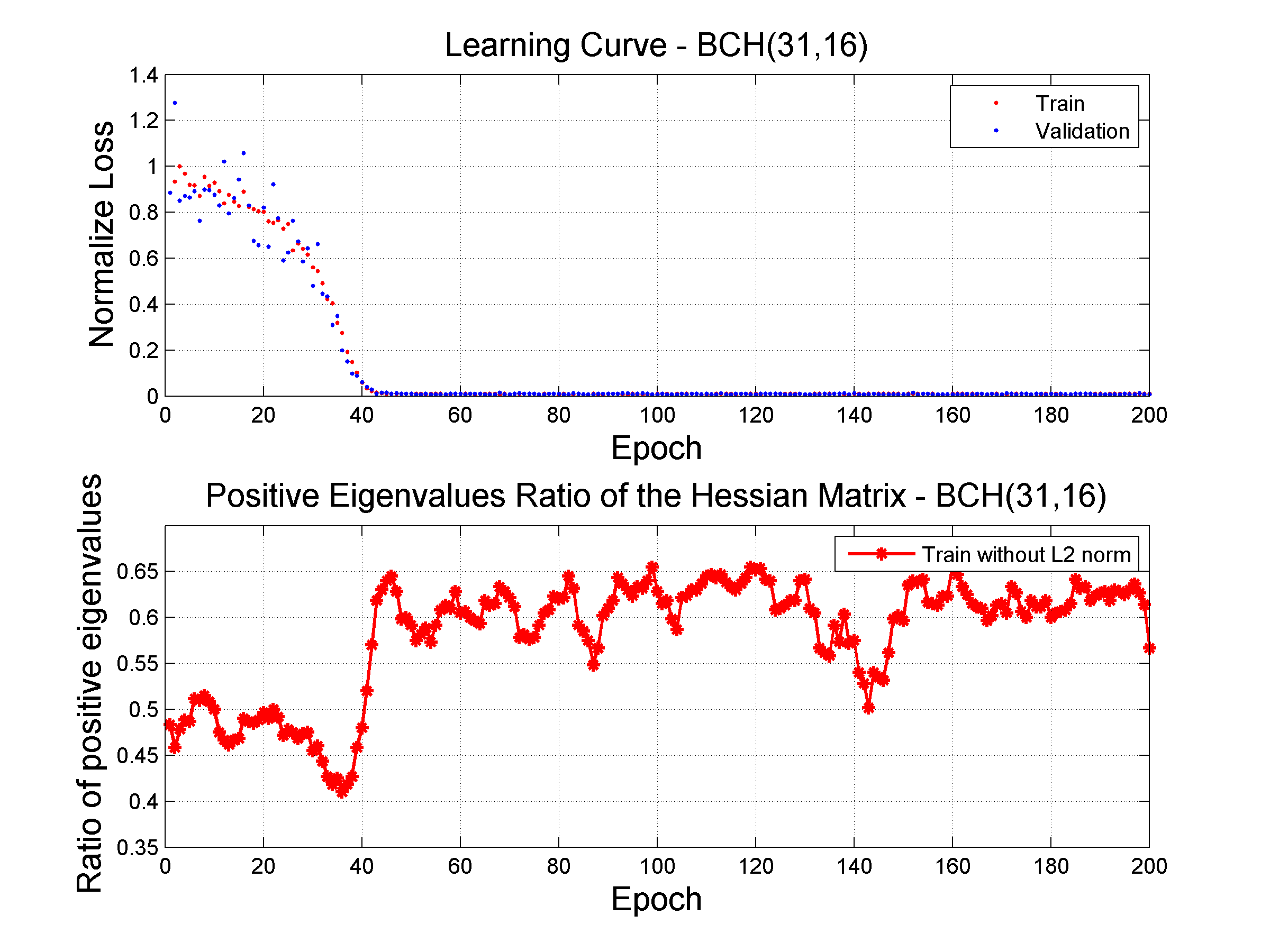}
	\caption{Learning Curve and Positive Eigenvalues Ratio of the Hessian Matrix for BCH(31,16) For Training without $l_2$-norm}
	\label{fig:bch_31_16_pr_nl2}
\end{figure}

\begin{figure}[thpb]
	\centering	\includegraphics[width=1.1\linewidth]{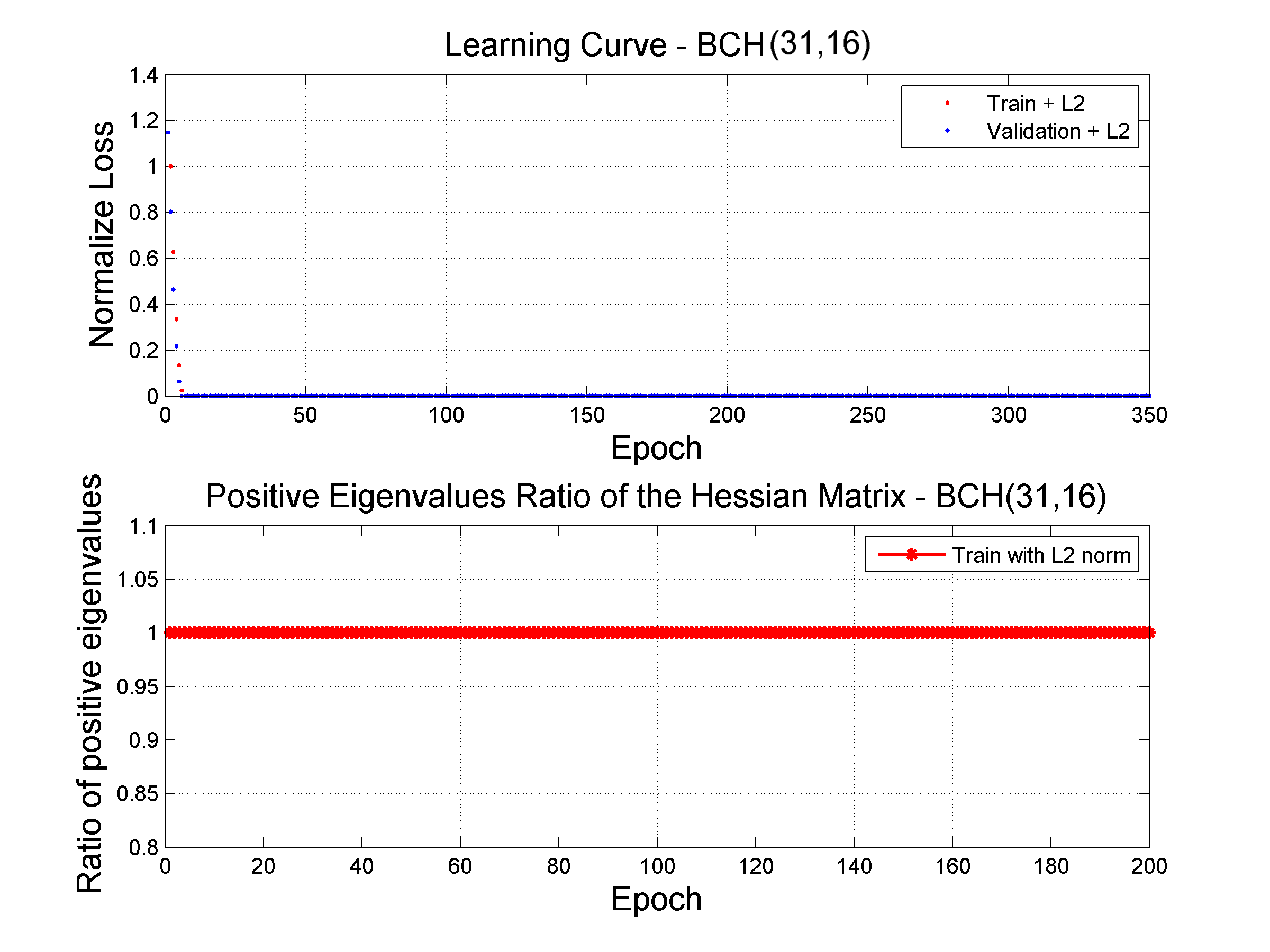}
	\caption{Learning Curve and Positive Eigenvalues Ratio of the Hessian Matrix for BCH(31,16) For Training with $l_2$-norm}
	\label{fig:bch_31_16_pr_wl2}
\end{figure}

To further investigate the phenomenon of the significant loss-drop and the $l_2$ acceleration, we computed the Hessian matrix of the deep neural decoder. Since the Hessian calculation demands high resources, we investigated the training process of a similar and smaller code, $BCH(31,16)$. As shown in Figures \ref{fig:bch_31_16_pr_nl2} and \ref{fig:bch_31_16_pr_wl2}, the training process of the $BCH(31,16)$ behaves in the same manner as the $BCH(63,45)$ code.

The Hessian matrix was evaluated during the training process. We calculated the condition number and the distribution of the eigenvalues of the Hessian matrix. The setting for the $BCH(31,16)$ code was: $ I_{permutations} = 10$ permutations, $I_{BP} =2$ BP-iterations, $c_{grad} = 0.1$, $\lambda = 100 $.

\begin{figure}[thpb]
	\centering	\includegraphics[width=1.1\linewidth]{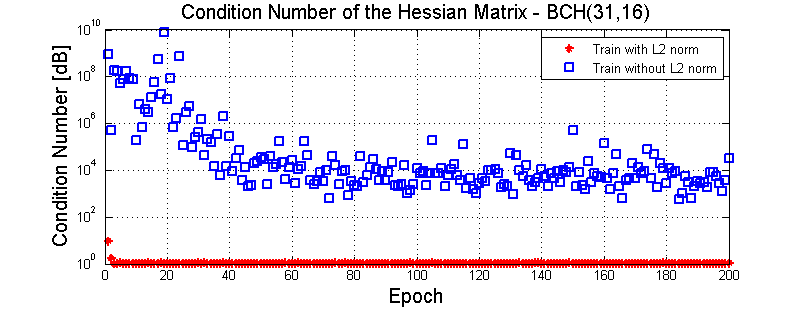}
	\caption{Condition Number of the Hessian Matrix during training for BCH(31,16) code}
	\label{fig:bch_31_16_cn}
\end{figure}

Figures \ref{fig:bch_31_16_pr_nl2} and \ref{fig:bch_31_16_cn} demonstrate the significant loss-drop properties. Whereas in epochs 1-20 the loss and the BER do not improve significantly and the positive eigenvalues ratio is low, epoch 20-40 serves as a turning point: the loss and the BER decrease rapidly and the positive eigenvalues ratio increases at the same time.

Figures \ref{fig:bch_31_16_pr_wl2} and \ref{fig:bch_31_16_cn} further stress this matter: the positive eigenvalues ratio is high right from the beginning, and accordingly the loss presents no initial-plateau to begin with. Put in other words, the Hessian rapidly becomes similar to a scaled identity matrix. The equivalent loss-surface is isotropic, which results in an accelerated learning process.

It is of no surprise that adding an $l_2$ term to the loss brings the Hessian closer to an identity matrix. Yet, the notable training acceleration and the performance improvement are a result of a gentle setting of parameters and the specific optimization problem discussed. 
\vspace{-5ex}

\vspace{-2ex}

%
%
%
%
%
%
\vspace{-2ex}
\bibliographystyle{IEEEtran}
\bibliography{IEEEexample}

\begin{thebibliography}{10}
\providecommand{\url}[1]{#1}
\csname url@samestyle\endcsname
\providecommand{\newblock}{\relax}
\providecommand{\bibinfo}[2]{#2}
\providecommand{\BIBentrySTDinterwordspacing}{\spaceskip=0pt\relax}
\providecommand{\BIBentryALTinterwordstretchfactor}{4}
\providecommand{\BIBentryALTinterwordspacing}{\spaceskip=\fontdimen2\font plus
\BIBentryALTinterwordstretchfactor\fontdimen3\font minus
  \fontdimen4\font\relax}
\providecommand{\BIBforeignlanguage}[2]{{%
\expandafter\ifx\csname l@#1\endcsname\relax
\typeout{** WARNING: IEEEtran.bst: No hyphenation pattern has been}%
\typeout{** loaded for the language `#1'. Using the pattern for}%
\typeout{** the default language instead.}%
\else
\language=\csname l@#1\endcsname
\fi
#2}}
\providecommand{\BIBdecl}{\relax}
\BIBdecl

\bibitem{farsad2017detection}
N.~Farsad and A.~Goldsmith, ``Detection algorithms for communication systems
  using deep learning,'' \emph{arXiv preprint arXiv:1705.08044}, 2017.

\bibitem{AutoencoderComm}
T.~J. O'Shea and J.~Hoydis, ``An introduction to machine learning
  communications systems,'' \emph{arXiv preprint arXiv:1702.00832}, 2017.

\bibitem{samuel2017deep}
N.~Samuel, T.~Diskin, and A.~Wiesel, ``Deep mimo detection,'' \emph{arXiv
  preprint arXiv:1706.01151}, 2017.

\bibitem{liang2017iterative}
F.~Liang, C.~Shen, and F.~Wu, ``An iterative bp-cnn architecture for channel
  decoding,'' \emph{arXiv preprint arXiv:1707.05697}, 2017.

\bibitem{dorner2017deep}
S.~Dorner, S.~Cammerer, J.~Hoydis, and S.~ten Brink, ``Deep learning-based
  communication over the air,'' \emph{arXiv preprint arXiv:1707.03384}, 2017.

\bibitem{Shengliang}
P.~Shengliang, J.~Hanyu, W.~Huaxia, and Y.~Yu-Dong, ``Deep learning and its
  applications in communications systems - modulation classification,''
  \emph{Submitted to IEEE Communications Magazine}, 2017.

\bibitem{power_ofdm}
Y.~Hao, Y.~L. Geoffrey, and F.~J. Biing-Hwang, ``Power of deep learning for
  channel estimation and signal detection in ofdm systems,'' \emph{arXiv
  preprint arXiv:1708.08514}, 2017.

\bibitem{Haowen_Lin}
H.~Sihao and L.~Haowen, ``Fully optical spacecraft communications: Implementing
  an omnidirectional pv-cell receiver and 8mb/s led visible light downlink with
  deep learning error correction,'' \emph{arXiv preprint arXiv:1709.03222},
  2017.

\bibitem{nachmani}
E.~Nachmani, Y.~Be'ery, and D.~Burshtein, ``Learning to decode linear codes
  using deep learning,'' in \emph{54'th Annual Allerton Conf. On Communication,
  Control and Computing}, September 2016, arXiv preprint arXiv:1607.04793.

\bibitem{nachmani2017rnn}
E.~Nachmani, E.~Marciano, D.~Burshtein, and Y.~Be'ery, ``Rnn decoding of linear
  block codes,'' \emph{arXiv preprint arXiv:1702.07560}, 2017.

\bibitem{lugosch}
L.~Lugosch and W.~J. Gross, ``Neural offset min-sum decoding,'' in \emph{2017
  IEEE International Symposium on Information Theory}, June 2017, arXiv
  preprint arXiv:1701.05931.

\bibitem{nachmani2017deep}
E.~Nachmani, E.~Marciano, L.~Lugosch, W.~J. Gross, D.~Burshtein, and Y.~Beery,
  ``Deep learning methods for improved decoding of linear codes,'' \emph{IEEE
  Journal of Selected Topics In Signal Processing}, Feb. 2018.

\bibitem{tenbrink}
T.~Gruber, S.~Cammerer, J.~Hoydis, and S.~ten Brink, ``On deep learning-based
  channel decoding,'' \emph{accepted for CISS 2017, arXiv preprint
  arXiv:1701.07738}, 2017.

\bibitem{cammerer2017scaling}
S.~Cammerer, T.~Gruber, J.~Hoydis, and S.~ten Brink, ``Scaling deep
  learning-based decoding of polar codes via partitioning,'' \emph{arXiv
  preprint arXiv:1702.06901}, 2017.

\bibitem{quantumCodes}
S.~Krastanov and L.~Jiang, ``Deep neural network probabilistic decoder for
  stabilizer codes,'' \emph{arXiv preprint arXiv:1705.09334}, 2017.

\bibitem{pless1978fj}
F.~J. MacWilliams and N.~J.~A. Sloane, \emph{The theory of error-correcting
  codes}.\hskip 1em plus 0.5em minus 0.4em\relax Elsevier, 1977.

\bibitem{dimnik2009improved}
I.~Dimnik and Y.~Be'ery, ``Improved random redundant iterative hdpc decoding,''
  \emph{IEEE Transactions on Communications}, vol.~57, no.~7, pp. 1982--1985,
  2009.

\bibitem{rmsprop}
T.~Tieleman and G.~Hinton, ``Lecture 6.5-rmsprop: Divide the gradient by a
  running average of its recent magnitude,'' \emph{COURSERA: Neural Networks
  for Machine Learning}, vol.~4, no.~2, 2012.

\bibitem{cycle_reduce}
T.~R. Halford and K.~M. Chugg, ``Random redundant soft-in soft-out decoding of
  linear block codes,'' in \emph{Information Theory, 2006 IEEE International
  Symposium on}.\hskip 1em plus 0.5em minus 0.4em\relax IEEE, 2006, pp.
  2230--2234.

\bibitem{automorphism}
C.-C. Lu and L.~R. Welch, ``On automorphism groups of binary primitive bch
  codes,'' in \emph{Proc. IEEE Symposium on Information Theory}, June 1994, p.
  1951.

\bibitem{fossOSD}
M.~P. Fossorier and S.~Lin, ``Soft-decision decoding of linear block codes
  based on ordered statistics,'' \emph{IEEE Transactions on Information
  Theory}, vol.~41, no.~5, pp. 1379--1396, 1995.

\bibitem{dauphin2014identifying}
Y.~N. Dauphin, R.~Pascanu, C.~Gulcehre, K.~Cho, S.~Ganguli, and Y.~Bengio,
  ``Identifying and attacking the saddle point problem in high-dimensional
  non-convex optimization,'' in \emph{Advances in neural information processing
  systems}, 2014, pp. 2933--2941.

\end{thebibliography}


\end{document}